\documentclass[amsmath,trackchanges,default]{aastex702}

\usepackage{longtable} 
\usepackage{booktabs} 
\usepackage{tabularx} 
\usepackage{makecell} 
\usepackage{array} 
\usepackage{ragged2e} \newcolumntype{Y}{>{\RaggedRight\arraybackslash}X}
\newcolumntype{L}[1]{>{\RaggedRight\arraybackslash}p{#1}}

\begin{document}

\title{How Small is Large Enough? Determining Minimal Cluster Sizes for Molecule Adsorption on Interstellar Amorphous Ice}

\author[orcid=0000-0002-3360-3196,sname='Neyts']{Erik C. Neyts}
\affiliation{University of Antwerp, Department of Chemistry, Research group MOSAIC, Groenenborgerlaan 171, 2020 Antwerp, Belgium}
\email[show]{erik.neyts@uantwerpen.be} 
\correspondingauthor{Erik C. Neyts}

\author{Christopher King} 
\affiliation{University of Antwerp, Department of Chemistry, Research group MOSAIC, Groenenborgerlaan 171, 2020 Antwerp, Belgium}
\affiliation{University of Montana, Department of Chemistry, Chemistry Building 115, 32 Campus Drive, Missoula, MT 59812, USA}
\email{christopher.king@uantwerpen.be}

\author[orcid=0000-0003-3632-0754,sname='Grubova']{Irina Grubova} 
\affiliation{University of Antwerp, Department of Chemistry, Research group MOSAIC, Groenenborgerlaan 171, 2020 Antwerp, Belgium}
\email{Irina.Grubova@uantwerpen.be}

\author[orcid=0000-0002-4724-2782,sname='Vorsselmans']{Tobe Vorsselmans} 
\affiliation{University of Antwerp, Department of Chemistry, Research group MOSAIC, Groenenborgerlaan 171, 2020 Antwerp, Belgium}
\email{Tobe.Vorsselmans@uantwerpen.be}

\begin{abstract}
Binding energies of molecules on ice mantles are important to understand the evolution of molecular complexity in molecular clouds. They are often computed using density functional theory (DFT) calculations, typically on either small amorphous ice clusters or crystalline slabs. Since these calculations require an accurate description of the electronic structure, hybrid functionals with dispersion corrections, basis set superposition error corrections and zero point energy corrections are typically employed. This, however, comes at a high computational cost, so most often small ice clusters are considered, frequently containing no more than twenty water molecules or so. While several recent studies have explored binding energy distributions, the effect of finite cluster size remains insufficiently quantified. To address this gap, we perform DFT calculations using six different functionals, on ice clusters containing 10 to 100 H$_2$O molecules, separating the direct electronic effect of cluster truncation from the geometry-relaxation effects. As probe molecules, we use CO, CO$_2$ and NH$_3$. These calculations demonstrate that, irrespective of the molecule and functional used, interaction energies only start to converge from thirty to forty water molecules onwards. The dispersion energy flattens out earlier, whereas induction and polarisation effects require larger clusters to stabilise, particularly at structurally confined (cavity) sites. We conclude that ice cluster sizes of at least 30–40 water molecules are needed to obtain reliable binding energies, and that cluster size is as important as the choice of the functional.
\end{abstract}


\section{Introduction}\label{Introduction}
The interaction of atoms with interstellar dust grains and the formation of volatile species on their surfaces is a cornerstone of cold interstellar chemistry. In the cold ($< 20$ K), dense regions of the interstellar medium (ISM) and the midplanes of protoplanetary disks, sub-micrometer-sized refractory dust grains act as cold-traps, accumulating layers of 'dirty' molecular ice mantles primarily composed of water (H$_2$O), carbon monoxide (CO), carbon dioxide (CO$_2$), ammonia (NH$_3$), and other molecules \citep{boogertObservationsIcyUniverse2015,mcclureIceAgeJWST2023}. The binding energy ($E_b$) of an adsorbate on these ice surfaces is the fundamental thermodynamic quantity that dictates whether a molecule remains frozen in the solid phase or desorbs into the gas phase. Together with the local gas density and temperature, $E_b$ is therefore the primary physical parameter determining the location of snow lines in protoplanetary disks, where the thermal conditions of the nascent stellar system trigger the freeze-out of specific volatile species \citep{obergEFFECTSSNOWLINESPLANETARY2011}. Because snow lines govern the volatile and organic composition of assembling planetesimals and gas-giant atmospheres, accurate determinations of these boundaries are important for interpreting modern high-resolution astronomical observations, such as those from ALMA and the James Webb Space Telescope (JWST) \citep{mcclureIceAgeJWST2023,qiImagingCOSnow2013a}.

Beyond static spatial features like snow lines, binding energies are also important input parameters for time-dependent astrochemical gas-grain kinetic models. These networks simulate the chemical evolution of molecular clouds and star-forming regions over millions of years by solving a large set of coupled differential rate equations \citep{hasegawaModelsGasgrainChemistry1992}. Within these models, the thermal desorption rate constant ($k_{\textrm{des}}$) is typically expressed via a Polanyi-Wigner or Arrhenius-type relation \citep{fraserThermalDesorptionWater2001}. Because of the exponential dependence of $k_{\textrm{des}}$ on the binding energy, even a minor discrepancy of $1 - 2$ kcal/mol (equivalent to roughly 500-1000 K) in $E_b$ can alter predicted desorption rates and gas-to-ice abundance ratios by several orders of magnitude at typical dark molecular cloud temperatures. Furthermore, surface diffusion barriers ($E_{\textrm{diff}}$), which govern the rates of Langmuir-Hinshelwood surface reactions that synthesize complex organic molecules on grain surfaces, are routinely estimated in modeling networks as a direct fraction of the binding energy (typically $E_{\textrm{diff}} \approx 0.3 - 0.8 E_b$) \citep{hasegawaModelsGasgrainChemistry1992,cuppenGrainSurfaceModels2017}. Thus, uncertainties in calculated $E_b$ values cascade directly into the predicted chemical pathways.

Experimentally, binding energies of stable astrochemically relevant molecules are predominantly derived from temperature-programmed desorption (TPD) mass spectrometry \citep{collingsLaboratorySurveyThermal2004}. While TPD provides invaluable macroscopic measurements, laboratory conditions often struggle to fully decouple the intrinsic site heterogeneity of amorphous ice surfaces. Moreover, experimental measurements are highly sensitive to sub-monolayer surface coverage, ice morphology (e.g., porous versus compact amorphous solid water), and thermal history \citep{hornekaerInfluenceSurfaceMorphology2005,nobleThermalDesorptionCharacteristics2012,cuppenLaboratoryComputationalStudies2024}. To complement and interpret these experiments, quantum chemical modeling, and specifically density functional theory (DFT), has become a widespread tool.

However, obtaining accurate and reliable binding energies from DFT involves addressing several computational trade-offs. Foremost is the chosen level of theory, including the exact exchange-correlation functional and the atomic basis set. Standard local or semi-local functionals fail to describe long-range dispersion (van der Waals) forces, which are an important binding mechanism for physisorbed molecules like CO and CO$_2$ on ice surfaces. To remedy this, modern calculations incorporate dispersion-correction schemes (such as Grimme’s DFT-D3/D4) or non-local van der Waals density functionals \citep{grimmeDensityFunctionalTheory2011}.

An equally critical, yet far less systematically investigated factor is the structural representation of the underlying ice substrate. Computational studies typically choose between periodic boundary condition (PBC) models representing infinite crystalline or amorphous surfaces, and finite cluster models \citep{zamirriQuantumMechanicalInvestigations2019}. While PBC calculations naturally capture bulk properties and long-range electrostatic screening, cluster models offer significant practical advantages: they are highly flexible, allow for the localized evaluation of heterogeneous binding sites on complex amorphous surfaces without artificial periodic self-interaction, and permit the use of highly accurate wave function-based correlation methods (such as coupled cluster theory) to benchmark DFT results. In Table \ref{Table1}, an overview is presented of the ice models used in the recent literature.

Results coming forth from a cluster model, however, depend heavily on its size. Because dispersion and electrostatic interactions are long-range phenomena, they decay gradually with distance from the adsorption site. If the chosen cluster model is too large, the calculations become prohibitively heavy. If the chosen cluster model is too small, it fails to capture the cooperative hydrogen-bonding network of the ice substrate and artificially truncates the long-range dispersion tail, leading to severe finite-size artefacts and unphysical binding energies. Therefore, multiscale frameworks and machine-learning approaches have recently begun establishing embedding radii and benchmarking functional performance on small clusters (e.g.,\citet{groyneRobustBindingEnergy2025,bovolentaCOAdsorptionSites2025}). However, a systematic benchmark evaluating numerical binding energy convergence across standalone DFT ice clusters remains lacking in the astrochemistry literature.

To address this gap, this study systematically investigates the convergence behaviour of molecular interaction energies as a function of the underlying ice cluster size. We target three astrochemically relevant species representing different dominant interaction regimes: CO (weakly bound, dispersion-driven), CO$_2$ (quadrupolar, moderately bound), and NH$_3$ (strongly hydrogen-bonded). Interaction energies are evaluated on two distinct types of sites (dangling-H and oxygen-rich pockets) across three independent amorphous 100-molecule amorphous ice clusters. By testing six density functionals with diverse treatments of non-covalent interactions, we map the spatial and size requirements necessary to yield computationally robust interaction energies.

\startlongtable
\label{Table1}
\begin{deluxetable}{>{\raggedright\arraybackslash}p{3.2cm} 
                    >{\raggedright\arraybackslash}p{3.2cm} 
                    >{\raggedright\arraybackslash}p{4.5cm} 
                    >{\raggedright\arraybackslash}p{4.5cm}}
\tabletypesize{\tiny}
\tablecaption{Overview of recent DFT studies examining the interaction of molecules with water ice clusters. $BS$ stands for basis set; cryst. stands for crystalline. \label{tab:ice_models_summary}}
\tablehead{
\colhead{Author (Year)} & \colhead{Ice Model and Size} &
\colhead{Functional / Basis Set} & \colhead{Molecules Studied}
}
\startdata
        \citet{lambertsInterstellarCarbonMonosulfide2018} & DFT-cluster \newline 18--33 H$_2$O & MPWB1K; B3LYP; PBEh-3c \newline \textit{BS:} def2-TZVP; def2-mSVP & CS; HCS; H$_2$CS; CH$_3$S; CH$_3$SH \\
        \midrule
        \citet{ferreroBindingEnergiesInterstellar2020} & DFT-periodic; ONIOM \newline 12 layers (cryst., 24 H$_2$O); \newline 60 H$_2$O (ASW)& DFT: B3LYP-D3(BJ); M06-2X\newline low level: HF3c \newline \textit{BS:} A-VTZ* & H$_2$; N$_2$; O$_2$; HCl; CO; CO$_2$; OCS; HCN; CH$_4$; NH$_3$; H$_2$S; H$_2$O; HCONH$_2$; H$_2$CO; HCOOH; CH$_3$OH; CH$_3$CN; OH; NH$_2$; HCO; CH$_3$ \\
        \midrule
        \citet{duflotTheoreticalDeterminationBinding2021} & ONIOM \newline $\sim$24:140 H$_2$O & DFT: M06-2X; $\omega$B97X-D \newline low level: PM6; PM7R8 \newline \textit{BS:} 6-31+G**; def2-TZVP & H; C; N; O; NH; OH; H$_2$O; CH$_3$; NH$_3$ \\
        \midrule
        \citet{sameeraCH3RadicalBinding2021} & ONIOM \newline $\sim$50:115 H$_2$O & DFT: $\omega$B97X-D \newline low level: AMBER; AMOEB09 \newline \textit{BS:} def2-TZVP & CH$_3$O \\
        \midrule
        \citet{rimolaInteractionHCOCations2021} & DFT-cluster \newline 14--24 H$_2$O & BHLYP \newline \textit{BS:} 6-31(1+,3+)G(d,p) & HCO$^+$; Solvated electron \\
        \midrule
        \citet{bovolentaBindingEnergyEvaluation2022} & DFT-cluster \newline 22 H$_2$O & PBE-D3BJ; B97-2-D3BJ; CAM-B3LYP-D3BJ; TPSSH-D3BJ; MPWB1K-D3BJ; M05-D3BJ \newline \textit{BS:} def2-TZVP & H$_2$; N$_2$; CH$_4$; CH$_3$; CO; CO$_2$; C$_2$H$_2$; NH$_3$; NHCH$_2$; NH$_2$; CH$_3$O; CH$_3$OH; CH$_2$OH; H$_2$S; H$_2$CO; H$_2$O; HCOOH; HF; HCN; HNC \\
        \midrule
        \citet{Enrique2022Quantum} & DFT-cluster \newline 18--33 H$_2$O & BHLYP-D3(BJ) \newline \textit{BS:} 6-31+G(d,p); 6-311++G(2df,2pd) & CH$_3$; NH$_2$; HCO; CH$_3$O; CH$_2$OH; NH; OH \\
        \midrule
        \citet{perreroBindingEnergiesInterstellar2022} & DFT-periodic \newline 12 layers (cryst., 24 H$_2$O); \newline 60 H$_2$O (ASW) & B3LYP-D3(BJ); M06-2X; HF3c \newline \textit{BS:} A-VTZ* & H$_2$S; H$_2$S$_2$; CS; CH$_3$SH; SO$_2$; OCS; H$_2$CS; C$_3$S; NS; HS; HS$_2$; HCS; SO; S$_2$; C$_4$S; C$_2$S; S \\
        \midrule
        \citet{piacentinoComputationalEstimationBinding2022} & DFT-cluster \newline 1--3 H$_2$O & M06-2X \newline \textit{BS:} aug-cc-pVDZ & PO; PO$_2$; PO$_3$; HPO; HPO$_2$; OPO; POOH; PO$_2$OH; H$_2$PO; PH$_3$; N$_2$; CO; CH$_4$; CO$_2$; C$_2$H$_4$; C$_2$H$_6$; H$_2$S; C$_2$H$_2$; H$_2$CO; C$_3$H$_8$; C$_3$H$_6$; CH$_3$NH$_2$; CH$_2$CCH$_2$; CH$_3$CCH; HCl; CH$_3$OH; NH$_3$; CH$_3$NC; H$_2$O; CH$_3$CN \\
        \midrule
        \citet{molpeceresCrackingPuzzleCO22023} & DFT-cluster \newline 33 H$_2$O (33 CO) & MN15-D3BJ \newline \textit{BS:} 6-31+G(d,p) & CO; OH; HCO$_2$; CO2 \\
        \midrule
        \citet{bovolentaInDepthExplorationCatalytic2024} & DFT-cluster \newline 22 H$_2$O; 64 H$_2$O pore & B97M-D3BJ; BMK; BHandHLYP-D3BJ; BHandHLYP-D4 \newline \textit{BS:} def2-TZVP; def2-SVP & NH$_3$; H$_2$CO; AMEOH \\
        \midrule
        \citet{perreroBindingEnergiesEthanol2024} & DFT-periodic \newline 12 layers (cryst.); \newline 60 H$_2$O (ASW) & B3LYP-D3(BJ); HF-3c \newline \textit{BS:} A-VTZ* & CH$_3$CH$_2$OH; CH$_3$CH$_2$NH$_2$ \\
        \midrule
        \citet{silAssessingRealisticBinding2024} & DFT-cluster \newline 20 H$_2$O & $\omega$B97X-D \newline \textit{BS:} 6-311+G(d,p) & CH; NH; OH; SH; CN; NS; NO \\
        \midrule
        \citet{benedettiCODiffusionInterstellar2026} & DFT-cluster \newline 22 H$_2$O & M06-2X-D3; MPWB1K-D3BJ \newline \textit{BS:} def2-TZVP & CO \\
        \midrule
        \citet{bovolentaCOAdsorptionSites2025} & DFT-cluster 22--60 H$_2$O \newline MLP (periodic ASW) 500 H$_2$O & MPWB1K-D3BJ-gCP \newline \textit{BS:} def2-TZVP & CO \\
        \midrule
        \citet{tieppoBuildingFormamideNsubstituted2025} & DFT-cluster \newline 1--3 H$_2$O; \newline 24--48 H$_2$O (cryst.) & M062X; MPWB1K \newline \textit{BS:} aug-cc-pVTZ/QZ; 6-311++G(3df,2p) & HNCO; H; H$_2$NCO; HNCHO; NH$_2$CHO \\
        \midrule
        \citet{kakkarBindingEnergiesInterstellar2025} & DFT-periodic \newline 12 layers (cryst., 24 H$_2$O); \newline 60 H$_2$O (ASW) & B3LYP-D3(BJ); M06-2X; HF3c \newline \textit{BS:} A-VTZ* & H$_2$CO; CH$_3$NH$_2$; CH$_3$OH; CH$_3$CN; CH$_3$CHO; c-CH$_2$CH$_2$O; NH$_2$CHO; HCOOH; CH$_3$CH$_2$OH; CH$_3$OCH$_3$; CH$_3$CH$_2$CN; CH$_3$NCO; CH$_3$COCH$_3$; CH$_3$CH$_2$CHO; CH$_3$COOH; HCOOCH$_3$; HOCH$_2$CHO; HOCH$_2$CH$_2$OH; CH$_3$COOCH$_3$ \\
        \midrule
        \citet{digenovaHotSulfurRocks2025} & DFT-cluster \newline 18 H$_2$O & $\omega$B97XD \newline \textit{BS:} aug-cc-pV(T+d)Z; ma-def2-TZVP & S($^1$D); H$_2$O; H$_2$OS; HOSH; H$_2$SO \\
        \midrule
        \citet{groyneRobustBindingEnergy2025} & ONIOM \newline $\sim$25:250 H$_2$O & B3LYP-D3(BJ):GFN2-xtb \newline \textit{BS}: 6-311+G(d,p) & NH$_3$; CO; CH$_4$ \\
        \midrule
        \citet{vorsselmansBindingEnergiesSmall2025b} & DFT-cluster \newline 33 H$_2$O & PBE0-D3(BJ) \newline \textit{BS:} 6-311++G(d,p) & CO; CH$_4$; NH$_3$; HCO$^+$ \\
        \midrule
        \citet{mozhegorovAtomAmorphousH22026} & DFT-cluster \newline 2--16 H$_2$O & MP2; DFT (various) \newline \textit{BS:} aug-cc-pVTZ & H \\
        \midrule
        \citet{vorsselmansEffectIceCharging2026} & DFT-cluster \newline 30 H$_2$O & RevPBE38-D3(0); B3PW91-D3(BJ); $\omega$B97M-D4rev \newline \textit{BS:} def2-TZVPPD & S; SO$_2$; OCS; H$_2$S \\
\enddata
\end{deluxetable}

\section{Methodology}
This study evaluates the interaction energies of CO, NH$_3$, and CO$_2$ with pristine surfaces and structural cavities, modeled using three independent amorphous ice clusters. The baseline substrates consist of 100-H$_2$O pristine parent clusters and 97-H$_2$O cavity-containing parent clusters. To assess how the electronic structure method influences this size dependence, calculations were performed using six density functionals: $\omega$B97X-D4rev, $\omega$B97M-V, M06-D3(0), B3LYP-D4, PBE0-D4, and PBE-D4, with $\omega$B97X-D4rev serving as our reference method. The spatial range of the interaction is analysed by systematically reducing the ice cluster size, by progressively 'peeling' water molecules farthest from the adsorption site away from the parent structures. Ten cluster sizes (including the parent frameworks) are evaluated for each unique combination of molecule, adsorption site, parent framework, and density functional. To compile the dataset presented herein, a total of 5,406 quantum chemical calculations were performed using the ORCA 6.1.1 software \citep{neeseSoftwareUpdateORCA2025,neeseSHARKIntegralGeneration2023,Helmich2021improved}.

The procedure was as follows. First, the three target molecules (adsorbates) are geometrically optimised at the $\omega$B97X-D4rev/def2-TZVP level. These optimised static structures are then used for single-point energy calculations across the five other density functionals to evaluate functional dependence. All optimisations were carried out using the standard TightSCF and OPT keywords in ORCA, governing the SCF and geometry optimisation convergence, respectively.

Next, amorphous solid water (ASW) clusters are prepared to serve as adsorption surfaces. Pristine amorphous clusters containing 100 H$_2$O molecules are generated using Packmol v21.2 \citep{martinezPACKMOLPackageBuilding2009}. Because of their size, they are optimised using a progressive four-step ladder of increasing theory, viz. HF-3c (for initial relaxation), PBE-D4/def2-SVP, $\omega$B97X-D4rev/def2-SVP, and finally $\omega$B97X-D4rev/def2-TZVP.

The cavity clusters were created by removing three adjacent water molecules from the surface of these optimised pristine structures. To prevent immediate collapse during a subsequent optimisation step, a Ne atom is temporarily placed in the resulting void; this atom is naturally expelled during the geometry optimisation, leaving behind an intact, relaxed cavity.

Adsorption complexes are subsequently constructed by placing the pre-optimised molecules near the pristine cluster surfaces and inside the cavity structures. Every resulting parent complex is then fully geometrically optimised at the $\omega$B97X-D4rev/def2-TZVP level. 

To obtain accurate adsorption strengths, counterpoise corrections for the basis set superposition error (BSSE) are applied \citep{boysCalculationSmallMolecular1970}. Here, we explicitly distinguish between two terms: the \textit{binding energy} ($E_\mathrm{b}$) and the \textit{interaction energy} ($E_\mathrm{int}$). The \textit{binding energy} represents the thermodynamic strength of adsorption on a fully relaxed surface and is computed exclusively for the fully geometrically optimised parent complexes ($N = 100$ for pristine, $N = 97$ for cavity) as:

\begin{equation}
E_\mathrm{b} = -(E_{\textrm{complex}} - E_{\textrm{mol}} - E_{\textrm{ice}} - (E_{\textrm{ice,gc,bc}} - E_{\textrm{ice,gc}}) - (E_{\textrm{mol,gc,bc}} - E_{\textrm{mol,gc}}))
\end{equation}

where $E_{\textrm{complex}}$ is the geometry-optimised energy of the ice cluster with the molecule adsorbed (the 'complex'), $E_{\textrm{mol}}$ is the energy of the isolated, optimised molecule in the gas phase, and $E_{\textrm{ice}}$ is the energy of the isolated, optimised parent ice cluster. Further, $E_{\textrm{ice,gc,bc}}$ is the single-point energy of the ice cluster in the geometry of the complex and using the basis set of the complex, and $E_{\textrm{ice,gc}}$ is the single-point energy of the ice cluster in the geometry of the complex and using the basis set of the ice cluster. Similarly, $E_{\textrm{mol,gc,bc}}$ is the single-point energy of the molecule in the geometry of the complex and using the basis set of the complex, and $E_{\textrm{mol,gc}}$ is the single-point energy of the molecule in the geometry of the complex and using the basis set of the molecule. All energy values are reported as positive quantities, where a larger value implies a stronger interaction.

To systematically investigate finite-size effects and determine how the ice environment influences the interaction strength, sub-clusters of size (H$_2$O)$_N$ are generated by sequentially truncating the outer solvation shells of each parent framework. This yields structures of $N = 10, 20, \cdots, 100$ H$_2$O molecules for the pristine parent clusters, and $N = 7, 17, \cdots, 97$ H$_2$O molecules for the cavity parent clusters. Importantly, the geometries of these truncated complexes are kept strictly frozen at the optimised coordinates of the parent adsorption complex. 

Because these peeled sub-clusters are not re-optimised, their resulting BSSE-corrected values reflect a static frozen-geometry state rather than a fully relaxed state. Consequently, these energies are strictly defined as \textit{interaction energies} ($E_\mathrm{int}$) rather than binding energies ($E_\mathrm{b}$). While this rigid fragment approach deliberately neglects structural relaxation (i.e., the reorganisation energy of the water network upon truncation), it provides two methodological advantages. First, it avoids artificial restructuring of the hydrogen-bonding network in the ASW that smaller clusters would undergo during unconstrained geometric relaxation. Second, it preserves the exact local geometry of the adsorption pocket across the entire truncation series from (H$_2$O)$_{7/10}$ up to (H$_2$O)$_{97/100}$. This isolates the pure, long-range electronic and electrostatic contributions of the surrounding ice matrix as a function of cluster size, free from the structural noise of nuclear relaxation.

Finally, to benchmark how different electronic structure methods describe the adsorption profile, this entire dataset across all parent and truncated complexes is evaluated using the five alternative density functionals. For the reference $\omega$B97X-D4rev functional, the values at $N = 100$ and $N = 97$ represent true binding energies ($E_\mathrm{b}$); for the five alternative functionals, these full-parent values represent interaction energies evaluated at the $\omega$B97X-D4rev geometry ($E_\mathrm{int}$). Across all functionals, all truncated sub-clusters ($N < 100/97$) yield strictly interaction energies ($E_\mathrm{int}$).

This suite ($\omega$B97X-D4rev, $\omega$B97M-V, M06-D3(0), B3LYP-D4, PBE0-D4, and PBE-D4) was selected to systematically evaluate the impact of different physical approximations on the modeled interaction profile. Spanning multiple rungs of Jacob's Ladder, the selection explores three key theoretical dimensions. First, we contrast the semiclassical, atom-pairwise dispersion corrections of Grimme \citep{caldeweyherExtensionD3Dispersion2017} (D3 and the charge-dependent D4) against the density-dependent non-local dispersion functional (VV10) featured in $\omega$B97M-V. Second, the influence of self-interaction errors and long-range charge transfer during adsorption is assessed by comparing global hybrids (PBE0, B3LYP, and M06) with range-separated hybrids ($\omega$B97X and $\omega$B97M). Third, we investigate the role of kinetic energy density meta-GGAs (M06 and $\omega$B97M), which inherently capture short-to-medium range correlation, versus standard GGA-based hybrids (PBE0, B3LYP, and $\omega$B97X) and a regular GGA functional (PBE). This diverse representation ensures that the benchmark is sensitive to both the electronic structure backbone and the treatment of long-range dispersion.

\section{Results}
\subsection{Convergence of Interaction Energies with Cluster Size}
\begin{figure*}[ht!]
\plotone{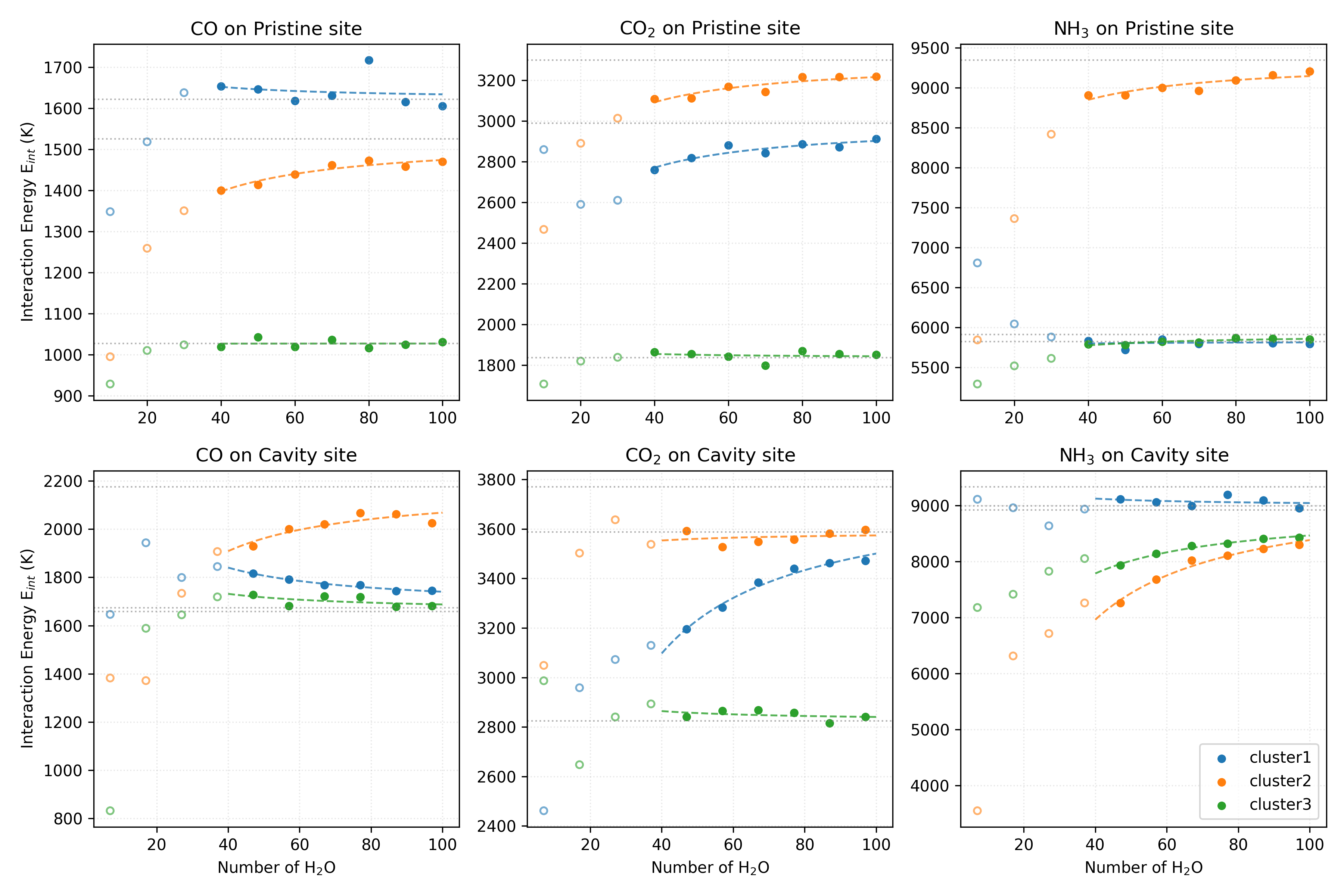}
\caption{Convergence of the interaction energy with respect to the number of H$_2$O molecules in the ice cluster, for the $\omega$B97X-D4rev functional, for each combination of cluster, molecule and adsorption site. The full symbols represent the data points used to generate the power law fit, depicted as the coloured dashed line. The grey horizontal lines represent the extrapolated interaction energy at infinite cluster size. This figure shows that there is a wide variation in interaction energies for the three clusters and that in most cases sizeable cluster sizes of about 40 H$_2$O molecules are needed to obtain reasonable interaction energies.}
\label{fig:fit_wrt_size}
\end{figure*}

To evaluate how the size of the ASW substrate influences interaction strengths, we systematically tracked the interaction energy ($E_{\textrm{int}}$) of CO, CO$_2$, and NH$_3$ as a function of the number of water molecules using the $\omega$B97X-D4rev functional. Figure \ref{fig:fit_wrt_size} displays these convergence profiles across three structurally independent ice clusters. 

The profiles demonstrate two prominent features. First, the wide variation in the absolute binding energies among the three individual clusters points to structural heterogeneity. For example, the pristine adsorption of CO at $N = 100$ ranges from approximately 1031 K on cluster 3 to over 1604 K on cluster 1. This spread highlights the highly amorphous and disordered nature of the generated ASW surfaces, which present a diverse topological landscape of binding environments.

Second, the curves also show a size-dependent stabilisation. For both pristine and cavity sites, the interaction energies exhibit rapid fluctuations at small cluster sizes ($N \leq 30$), represented as open circles, where the loss of even a few distant water molecules severely destabilises the local electrostatic field. Beyond $N \approx 40$ (represented as filled circles), the curves level off, with power-law fits (represented by the dashed lines) asymptotically approaching the infinite cluster size limit (represented by the horizontal grey lines). This indicates that a minimum of 40 water molecules is generally required to screen the adsorption site and yield a physically representative interaction energy.

\subsection{Convergence of the Dispersion Energy Component} \label{subsec:dispersion_convergence}
\begin{figure*}[ht!]
\plotone{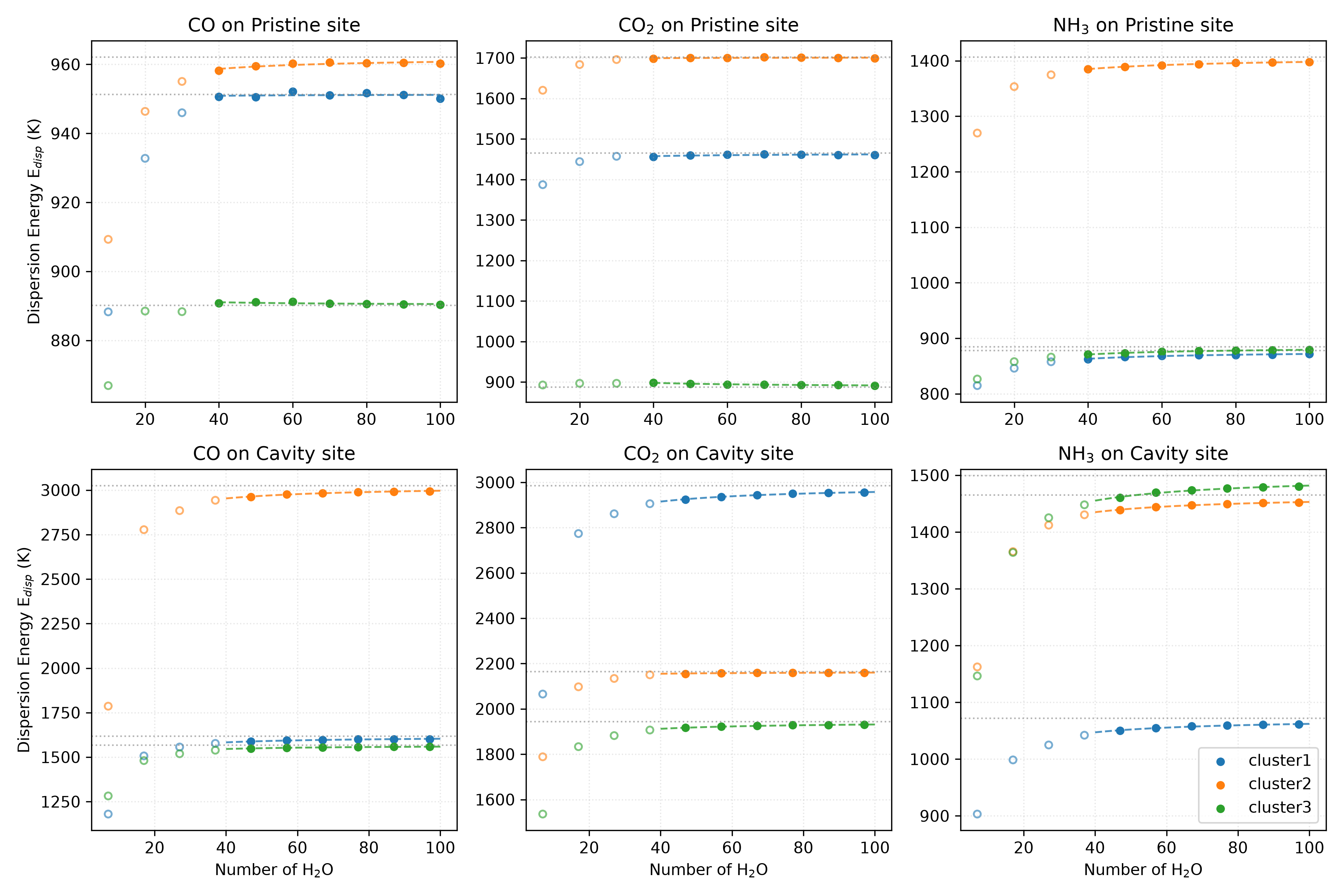}
\caption{Convergence of the dispersion energy with respect to the number of H$_2$O molecules in the ice cluster, for the $\omega$B97X-D4rev functional, for each combination of cluster, molecule and adsorption site. This figure shows that the dispersion energy converges quickly as a function of cluster size. Very similar results are obtained for the other functionals studied.}
\label{fig:fit_wrt_size_dispersion}
\end{figure*}

To isolate the physical drivers behind the cluster-size dependency, we analysed the convergence of the dispersion energy ($E_{\textrm{disp}}$) component. As shown in Figure \ref{fig:fit_wrt_size_dispersion} for the $\omega$B97X-D4rev functional, the dispersion contribution converges remarkably fast compared to the total interaction energy. For nearly all combinations of molecules and sites, $E_{\textrm{disp}}$ is essentially flatlined and fully converged by $N = 40$. This is equally true for all other functionals investigated, albeit to a somewhat lesser extent for the M06-D3(0) functional.

While volume integration over a bulk surface relaxes the dispersion scaling to $R^{-3}$, the finite size and concave curvature of a 100-molecule cluster restrict the number of long-range atom pairs. The interaction therefore retains a much steeper, more molecular-like decay closer to the microscopic $R^{-6}$ limit, making the adsorption energy heavily dependent on the local coordination environment (typically the first and second solvation shells). 

Consequently, stripping away water molecules that reside beyond a distance of approximately 6 to 8 \AA\ has negligible impact on the dispersion energy. This indicates that the lingering drift in $E_{\textrm{int}}$ observed at larger cluster sizes in Figure \ref{fig:fit_wrt_size} must arise from non-dispersive, longer-range electronic effects. Concretely, this slow convergence is driven by the cluster's long-range electric field, which takes more molecules to stabilise than dispersion, and the fact that adding distant water molecules to some degree alters the electron density of the water molecules at the adsorption site.

The distinction between short-range and long-range interactions is apparent when comparing the convergence profiles of the dispersion energy and the total interaction energy on the pristine sites. As shown in the dispersion profiles, $E_{\textrm{disp}}$ stabilises by $N = 40$ across all configurations and adsorbates, indicating that van der Waals interactions are primarily governed by the immediate, local coordination shell.

In contrast, the total interaction energy exhibits a configuration-dependent convergence lag that largely subsides by $N = 40-50$. As a first diagnostic, we calculate the local electrostatic potential at the adsorption site as a simple Coulomb sum over the ice-only atoms (see Tables \ref{tab:pristine_combined} and \ref{tab:cavity_combined}). This quantity is naturally weighted toward the immediate coordination environment rather than the cluster as a whole. On pristine surfaces, this local potential settles into a comparatively narrow, stable range by $N \approx 30-40$ for essentially all three adsorbates and all three parent clusters, mirroring the convergence window already reported for $E_{\textrm{int}}$ on these sites. CO$_2$ shows somewhat more persistent scatter than CO or NH$_3$ even at larger $N$, consistent with its known sensitivity to the local field gradient rather than the field itself, but the overall pattern on pristine surfaces is one of rapid stabilisation closely tracking the interaction-energy convergence.

Cavity sites show a markedly different picture. Here the local electrostatic potential continues to evolve well beyond $N = 40-50$ for several adsorbate/cluster combinations: for CO on cluster 2, it decays steadily in magnitude from –0.65 at $N = 17$ to –0.05 by $N = 97$ without ever plateauing early; for NH$_3$ on cluster 2, it climbs progressively and only stabilises around $N \approx 60$; and several CO$_2$ and NH$_3$ series retain visible drift out to $N = 77-87$ before narrowing. This slower stabilisation at cavity sites, relative to the comparatively fast convergence on pristine surfaces, indicates that the longer-range polarisation network is more extended — and takes longer to capture — when the adsorbate sits inside a more enclosed, higher-coordination pocket. Taken together, the site-specific electrostatic potential data show that the residual, non-dispersive convergence lag in $E_{\textrm{int}}$ tracks the slow stabilisation of the local electrostatic and induction environment at the adsorption site, and that this lag is itself site-dependent, converging faster on pristine surfaces than in cavities.

\begin{deluxetable*}{lrrrrrrrrr}
\tablewidth{0pt}
\tablecaption{Description of the local electrostatic potential (in V) at the adsorption site for CO, CO$_2$, and NH$_3$ on pristine ice clusters of increasing size. $N$ stands for the number of H$_2$O molecules in the cluster. \label{tab:pristine_combined}}
\tablehead{
\colhead{$N$} & \multicolumn{3}{c}{CO} & \multicolumn{3}{c}{CO$_2$} & \multicolumn{3}{c}{NH$_3$} \\
\cline{2-4} \cline{5-7} \cline{8-10}
\colhead{} & \colhead{Cluster 1} & \colhead{Cluster 2} & \colhead{Cluster 3} &
\colhead{Cluster 1} & \colhead{Cluster 2} & \colhead{Cluster 3} &
\colhead{Cluster 1} & \colhead{Cluster 2} & \colhead{Cluster 3}
}
\startdata
10  & $-$0.0126 & $-$0.2096 & $-$0.0518 & $-$0.2432 & $-$0.3455 & $-$0.0918 &  0.2314   &  0.0497   & 0.1902 \\
20  &  0.1019   & $-$0.4967 &  0.0756   & $-$0.1543 & $-$0.4838 & $-$0.0924 &  0.2320   & $-$0.0432 & 0.3696 \\
30  &  0.1635   & $-$0.4446 &  0.0630   &  0.0745   & $-$0.7147 & $-$0.1331 &  0.2284   & $-$0.0492 & 0.2896 \\
40  &  0.1180   & $-$0.4211 &  0.0318   & $-$0.2237 & $-$0.5975 & $-$0.1789 &  0.1733   &  0.0035   & 0.2686 \\
50  &  0.1659   & $-$0.4098 &  0.0785   & $-$0.0926 & $-$0.5716 & $-$0.1458 &  0.1389   & $-$0.0004 & 0.2705 \\
60  &  0.1437   & $-$0.4290 &  0.0364   & $-$0.0345 & $-$0.6418 & $-$0.1803 &  0.2182   &  0.0247   & 0.2709 \\
70  &  0.1531   & $-$0.4836 &  0.0830   & $-$0.0719 & $-$0.6422 & $-$0.1019 &  0.1699   & $-$0.0616 & 0.2476 \\
80  &  0.1510   & $-$0.4181 &  0.0651   &  0.0015   & $-$0.6914 & $-$0.1757 &  0.2478   &  0.0222   & 0.2961 \\
90  &  0.1828   & $-$0.4630 &  0.1094   & $-$0.0495 & $-$0.6562 & $-$0.1035 &  0.2109   & $-$0.0076 & 0.3169 \\
100 &  0.1966   & $-$0.4382 &  0.0984   & $-$0.0572 & $-$0.6198 & $-$0.0931 &  0.2148   &  0.0147   & 0.3013 \\
\enddata
\end{deluxetable*}

\begin{deluxetable*}{lrrrrrrrrr}
\tablewidth{0pt}
\tablecaption{Description of the local electrostatic potential (in V) at the adsorption site for CO, CO$_2$, and NH$_3$ on the cavity ice clusters of increasing size. $N$ stands for the number of H$_2$O molecules in the cluster. \label{tab:cavity_combined}}
\tablehead{
\colhead{$N$} & \multicolumn{3}{c}{CO} & \multicolumn{3}{c}{CO$_2$} & \multicolumn{3}{c}{NH$_3$} \\
\cline{2-4} \cline{5-7} \cline{8-10}
\colhead{} & \colhead{Cluster 1} & \colhead{Cluster 2} & \colhead{Cluster 3} &
\colhead{Cluster 1} & \colhead{Cluster 2} & \colhead{Cluster 3} &
\colhead{Cluster 1} & \colhead{Cluster 2} & \colhead{Cluster 3}
}
\startdata
7  & $-$0.1804 & $-$0.4731 & $-$0.0926 & $-$0.1999 &  $-$0.0870 & $-$0.2976 &  0.2062 & $-$0.1989 & 0.2882 \\
17  & $-$0.0101 & $-$0.6496 & $-$0.0034 & $-$0.1775 &   0.1776   & $-$0.3251 &  0.4733 &  0.3525   & 0.1374 \\
27  & $-$0.0851 & $-$0.3588 & $-$0.1595 & $-$0.3700  &   0.2066   & $-$0.2974 &  0.4225 &  0.3961   & 0.3122 \\
37  & $-$0.0726 & $-$0.2614 & $-$0.1746 & $-$0.1527  &   0.1569   & $-$0.2724 &  0.5543 &  0.4059   & 0.1259 \\
47  & $-$0.0149 & $-$0.2182 & $-$0.1472 & $-$0.0904  &   0.2814   & $-$0.2735 &  0.5148 &  0.4511   & 0.1883 \\
57  & $-$0.0150 & $-$0.1433 & $-$0.1166 & $-$0.2035  &   0.2175   & $-$0.2704 &  0.4700 &  0.5775   & 0.3316 \\
67  &  0.0577   & $-$0.2072 & $-$0.1255 & $-$0.1045  &   0.2545   & $-$0.3008 &  0.4824 &  0.5276   & 0.1935 \\
77  &  0.1100   & $-$0.0627 & $-$0.0698 & $-$0.0235  &   0.2476   & $-$0.2192 &  0.6348 &  0.5276   & 0.3330 \\
87  &  0.0762   & $-$0.0465 & $-$0.0677 & $-$0.0456  &   0.2990   & $-$0.1988 &  0.5635 &  0.5751   & 0.3494 \\
97 &  0.0573   & $-$0.0457 & $-$0.0167 & $-$0.0444  &   0.2807   & $-$0.1547 &  0.5679 &  0.5750   & 0.3877 \\
\enddata
\end{deluxetable*}

\subsection{Functional Independence of Cluster Size Convergence} \label{subsec:func_indep}
To determine whether the rate of convergence of the total interaction energy is sensitive to the quantum chemical method used, we compared the Mean Absolute Deviation (MAD) from the parent cluster baseline ($N = 100$ or $97$) across the six investigated density functionals. 

\begin{figure*}[ht!]
\plotone{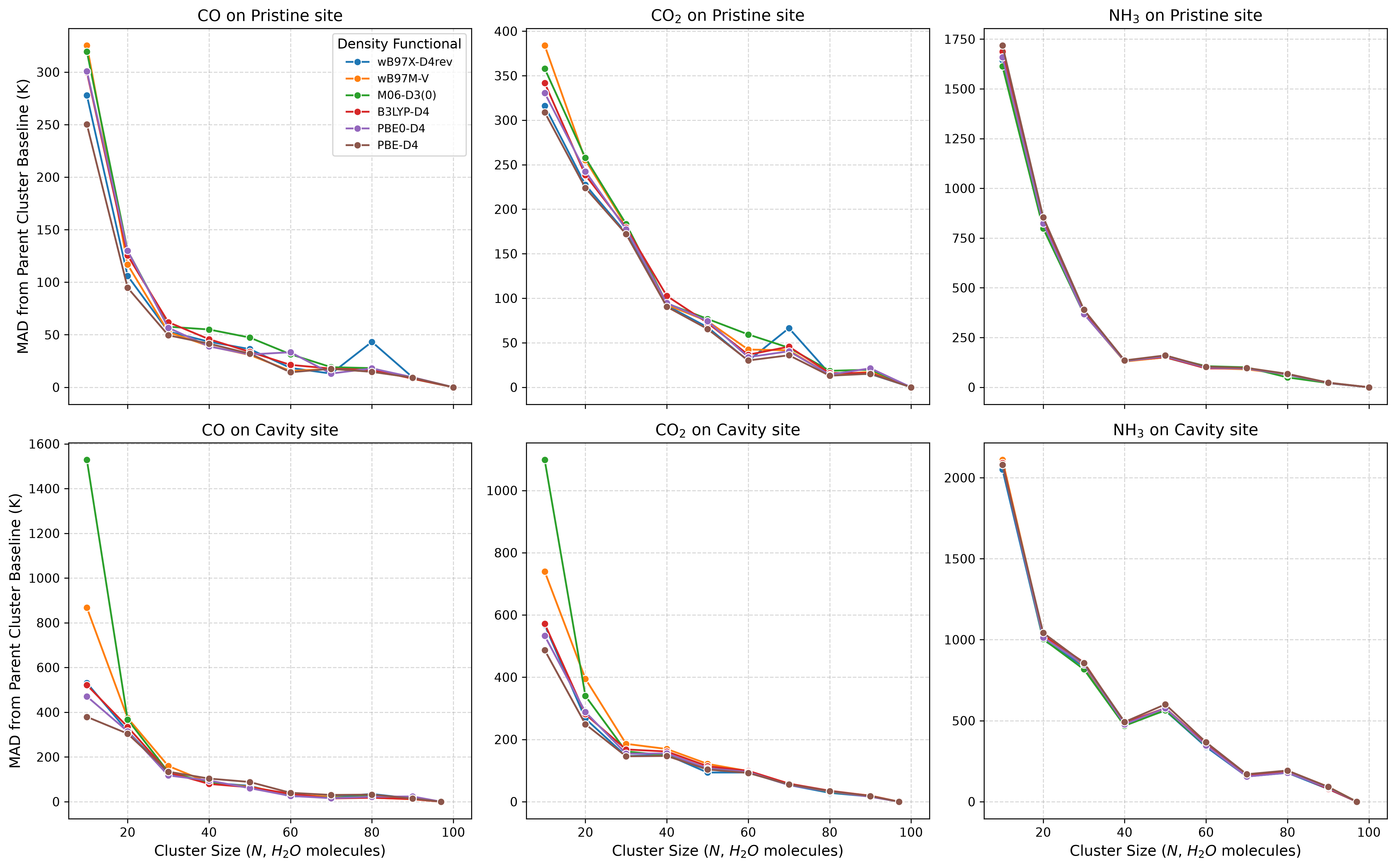}
\caption{Comparison of the convergence behaviour of the six functionals employed. Each curve for each functional represents the deviation of the interaction energy relative to the parent cluster ($N = 100$ for the pristine cluster, $N = 97$ for the cluster with the cavity site), averaged over the three clusters. This figure shows that the convergence with respect to size is essentially independent of the choice of functional.}
\label{fig:convergence_functionals}
\end{figure*}

As illustrated in Figure \ref{fig:convergence_functionals}, the MAD curves for all six functionals are virtually superimposed for every adsorbate and adsorption site. Whether employing global hybrids with semiclassical dispersion (like B3LYP-D4) or range-separated meta-GGAs with non-local dispersion (like $\omega$B97M-V), or even a standard GGA-functional but with charge-dependent dispersion (like PBE-D4), the physical decay of the interaction energy deviation as a function of $N$ remains identical. Thus, the figure shows that for each functional, convergence toward the $E_{\textrm{int}}$ value at $N = 100$ or $N = 97$ evolves nearly identically as function of cluster size. This demonstrates that cluster size convergence is a fundamental geometric and electrostatic property of the physical system, rather than an artefact of the chosen electronic structure method.

\subsection{Average Binding Energies: Pristine vs. Cavity Environments}
\begin{figure*}[ht!]
\plotone{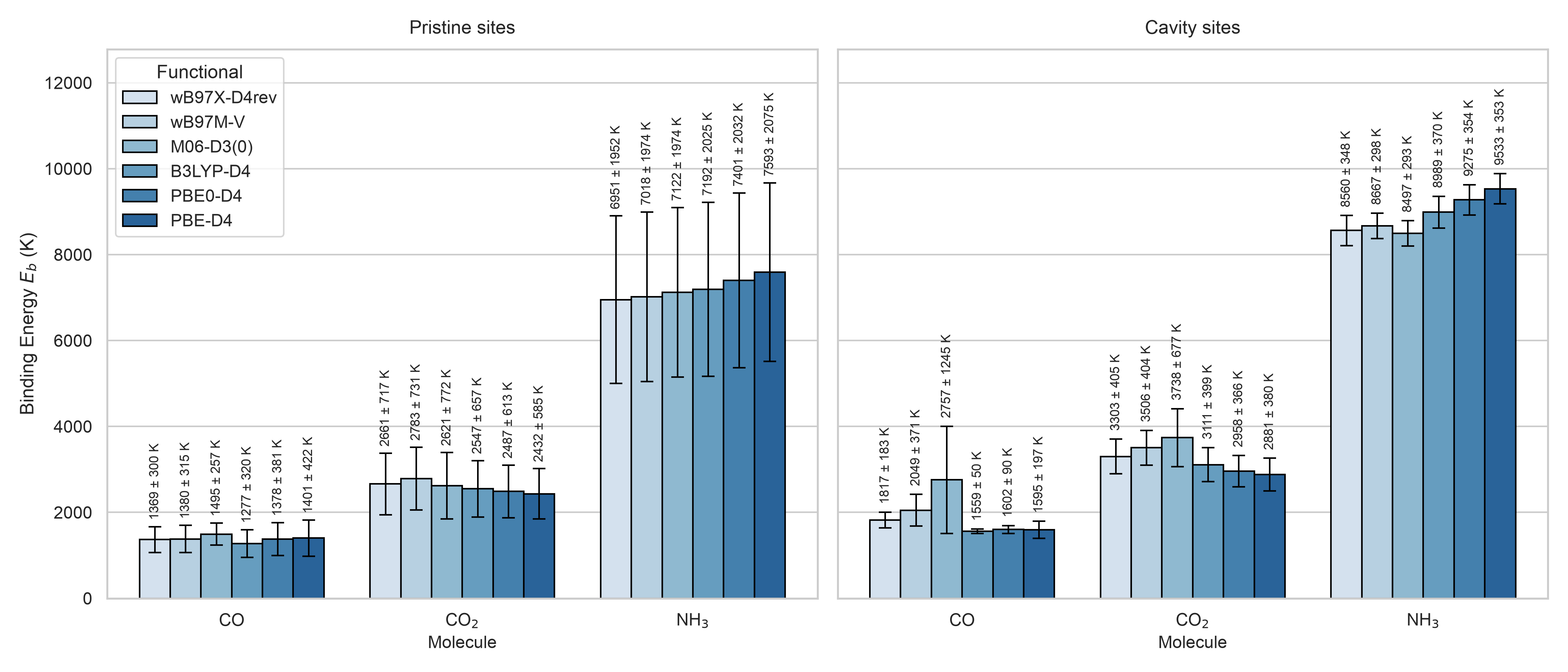}
\caption{Binding energies of CO, CO$_2$ and NH$_3$ on pristine and cavity-containing ice clusters consisting of 100 and 97 H$_2$O molecules, respectively. The error bars represent the spread in adsorption energies across the three clusters.}
\label{fig:BE100_functionals}
\end{figure*}

The average binding energies computed on the fully optimised parent clusters ($N = 100$ for pristine; $N = 97$ for cavity) using the $\omega$B97X-D4rev functional are summarised in Figure \ref{fig:BE100_functionals}. The figure shows two trends. First, adsorption inside structural cavities consistently yields higher binding energies than adsorption on pristine flat surfaces. Specifically, the average $E_b$, calculated with the $\omega$B97X-D4rev functional, increases from $1369 \pm 300$ K to $1817 \pm 183$ K for CO, from $2661 \pm 717$ K to $3303 \pm 405$ K for CO$_2$, and from $6951 \pm 1952$ K to $8560 \pm 348$ K for NH$_3$. This stabilisation is driven by the higher coordination number inside the cavity pocket, which allows the adsorbate to form multiple simultaneous interactions with the surrounding water molecules. Very similar changes are observed for the five other functionals.

Second, the standard deviation (represented by the error bars) is significantly smaller for cavity sites than for pristine surfaces. For example, the spread for NH$_3$ drops from $\pm 1952$ K on pristine sites to just $\pm 348$ K inside cavities. This indicates that while pristine surfaces offer a highly heterogeneous mix of weakly and strongly binding sites, structural cavities present a much more homogeneous, structurally constrained, and thermodynamic-favouring binding environment.

\subsection{Analysis of Dispersion Fractions Across Density Functionals} \label{subsec:analysis_dispersion}
\begin{figure*}[ht!]
\plotone{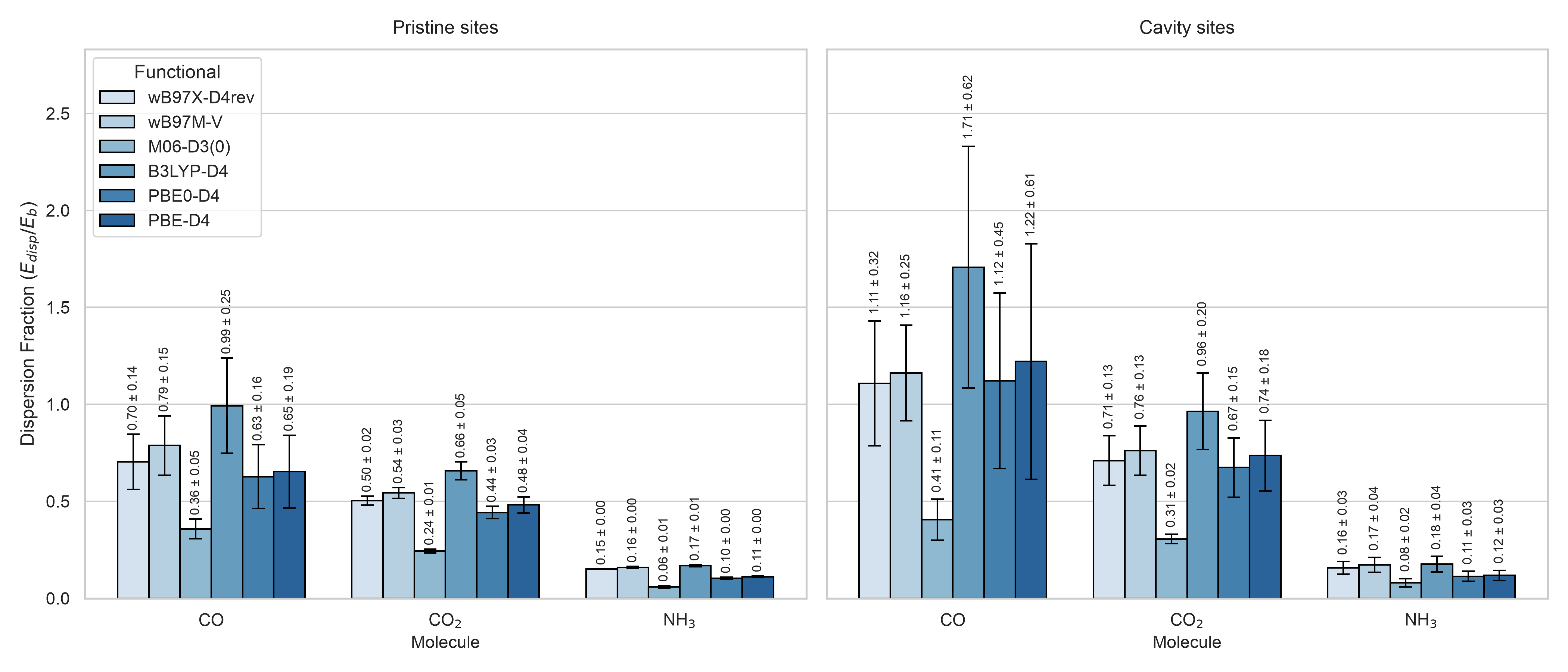}
\caption{Dispersion energy as fraction of the total binding energy per species, site and functional.}
\label{fig:BE100_functionals_dispersion}
\end{figure*}

Figure \ref{fig:BE100_functionals_dispersion} illustrates the dispersion energy expressed as a fraction of the total binding energy ($E_{\textrm{disp}}/E_b$) across the six density functionals for each adsorbate and binding site. While the total interaction energy profiles converge in a functional-independent manner (as shown in Section \ref{subsec:func_indep}), the absolute partitioning between dispersive and non-dispersive contributions varies significantly depending on the underlying electronic structure method.

A hierarchy emerges among the functionals, which remains consistent across all molecular species and adsorption environments. B3LYP-D4 consistently yields the highest dispersion fraction, reaching a maximum of $1.71 \pm 0.62$ for CO inside cavity sites and $0.99 \pm 0.25$ on pristine sites. This is followed by the pure GGA and global hybrid functionals PBE-D4 and PBE0-D4, which exhibit intermediate-high fractions (e.g., $1.22 \pm 0.61$ and $1.12 \pm 0.45$ for CO in cavities, respectively). The range-separated hybrid functionals, $\omega$B97M-V and $\omega$B97X-D4rev, present a consistent, moderate-to-high dispersion ratio (averaging $\approx 1.16$ and $\approx 1.11$ for CO in cavities). Conversely, the meta-GGA hybrid M06-D3(0) undercuts all other methods, exhibiting the lowest dispersion fraction, dropping to $0.41 \pm 0.11$ for CO in cavities, $0.24 \pm 0.01$ for CO$_2$ on pristine sites, and under $0.06 \pm 0.01$ for NH$_3$ on pristine sites.

The chemical nature of the adsorbate heavily modulates the magnitude of these fractions. For the weakly polar, dispersion-dominated CO molecule, the explicit dispersion component frequently equals or exceeds the net binding energy ($E_{\textrm{disp}}/E_b \ge 1.0$) for five out of the six functionals in cavity sites. For the quadrupolar CO$_2$ molecule, the dispersion fraction decreases but remains substantial, ranging from 0.24 to 0.96 depending on the functional and site. For the strongly polar, hydrogen-bonding NH$_3$ molecule, the fraction drops dramatically across all methods, never exceeding 0.18, as electrostatics and induction dominate the total interaction energy.

\subsection{Statistical Evaluation of Convergence Errors}
\begin{figure*}[ht!]
\plotone{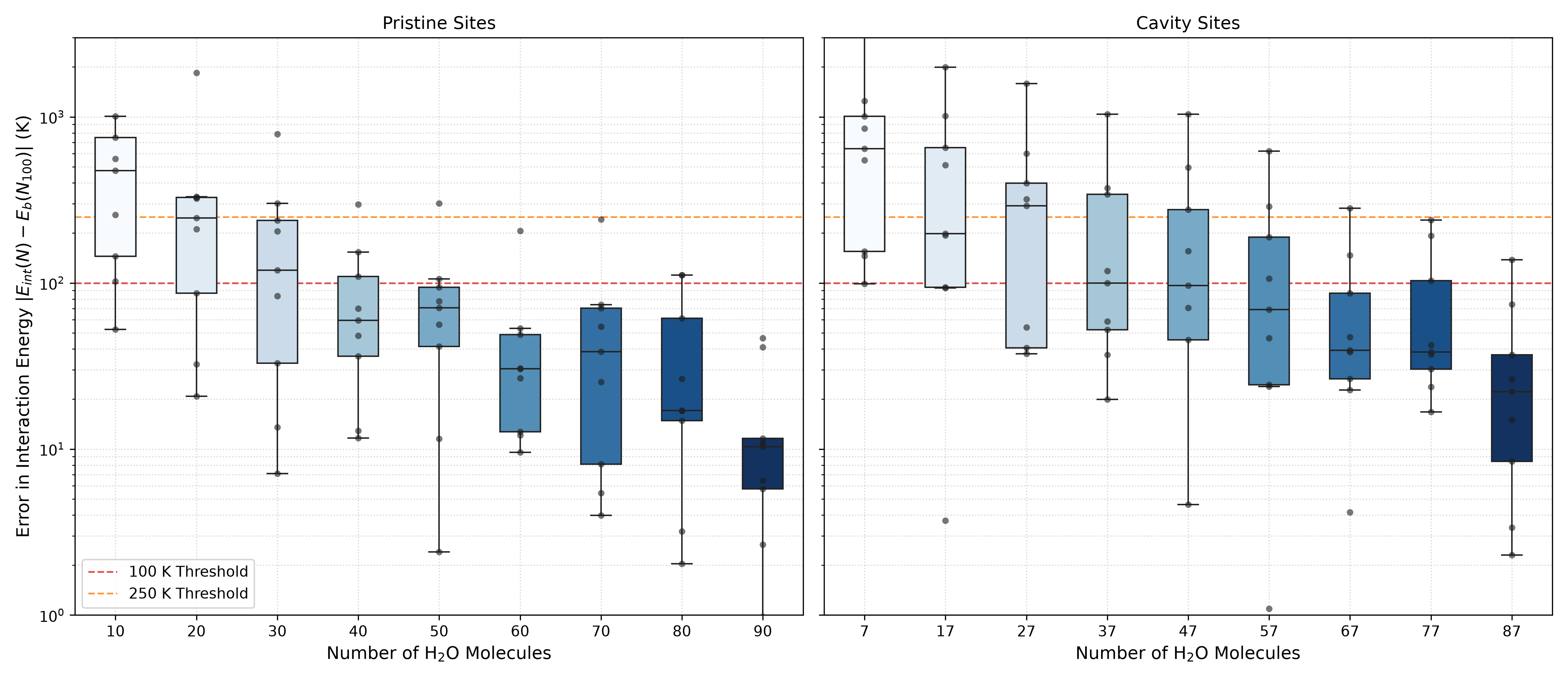}
\caption{Cumulative convergence for two convergence criteria: error in interaction energy of 100 K and 250 K with respect to the value for the 100 H$_2$O parent cluster. The figure shows that for both the pristine and the cavity sites, the median of the error drops below 100 K from 40 molecules.}
\label{fig:cumulative_convergence}
\end{figure*}

To provide a quantitative recommendation for computational models, Figure \ref{fig:cumulative_convergence} presents the boxplot distribution of the absolute convergence error, $\lvert E_{\textrm{int}}(N) - E_b(N_{\textrm{max}})\rvert$, across all sampled configurations for both pristine and cavity sites.

On pristine surfaces, a cluster size of $N \ge 40$ successfully reduces the median convergence error below the 100 K threshold (red dashed line), with the corresponding interquartile range falling entirely below this limit by $N = 50$. In comparison, the cavity environments display a significantly wider statistical dispersion across all sizes, reflecting the structural heterogeneity inherent to these localised defects. Although the median error for the cavity sites drops below the 100 K mark relatively early at $N = 37$, the overall error distribution remains broad. Specifically, the upper quartiles and several individual configurations continue to exceed the 250 K threshold (orange dashed line) up to $N = 57$.

As the cluster size increases, this statistical spread gradually contracts. By $N = 87$, the median error drops to approximately 25 K, and the vast majority of the sampled configurations fall well below the 100 K threshold, though a minor residual scatter remains slightly above it. These trends indicate that the geometric confinement and irregular coordination of cavity environments introduce highly variable, longer-range polarisation pathways that are sensitive to the specific amorphous arrangement. Consequently, while a smaller cluster size ($N \approx 40$) is often sufficient to capture the average binding behavior on flat surfaces, larger substrate models ($N \ge 77-87$) are necessary in disordered cavity environments to consistently minimize worst-case convergence errors across different configurations.

\section{Discussion}
\subsection{Physical Drivers of Convergence: Dispersion vs. Electrostatics and Induction}
The rapid convergence of the dispersion energy shown in Figure \ref{fig:fit_wrt_size_dispersion}, contrasted with the slower, fluctuating convergence of the total interaction energy in Figure \ref{fig:fit_wrt_size}, provides insight into the physical forces driving adsorption on ASW. 

The three target molecules represent distinct chemical regimes: CO is a weakly polar molecule with a very small dipole moment, meaning its adsorption is dominated by dispersion and localised induction. CO$_2$ is non-polar but possesses a large quadrupole moment and high polarisability, resulting in substantial dispersion and local electrostatic field-gradient interactions. NH$_3$ is highly polar, acting as a strong hydrogen-bond donor and acceptor, which induces a large local polarisation in the ice matrix. The adsorbate identity is thus central to how much of $E_{\textrm{int}}$ is dispersion vs. electrostatic (see section \ref{subsec:analysis_dispersion}). CO$_2$ specifically shows more persistent noise during convergence than CO or NH$_3$ even on pristine sites, consistent with its quadrupole/field-gradient character.

Because dispersion interactions on a molecular scale decay very rapidly with distance ($R^{-6}$), they are fully captured within the first 30 to 40 water molecules surrounding the adsorption site. Beyond this size, the addition of further water molecules does not alter the dispersion contribution. In contrast, electrostatic interactions (such as dipole-dipole, which decays as $R^{-3}$, and dipole-quadrupole, which decays as $R^{-4}$) and polarisation/induction forces are much longer-ranged. 

This effect is most pronounced for cavity sites. Indeed, while $N \approx 30-40$ is in most cases sufficient for both the dispersion and electrostatics to start to converge on pristine sites, $E_{\textrm{int}}$ requires larger clusters on cavity sites because electrostatics stay erratic while dispersion is flat. The site potential keeps drifting out to $N \approx 60-90$ in the cavity for multiple adsorbates (including CO, which is otherwise dispersion-dominated), while dispersion itself is already flat by $N = 40$ everywhere.

The stark differences in the dispersion fractions observed among the six density functionals are directly rooted in their theoretical formulation and how they partition non-covalent interactions along Jacob's Ladder. The high dispersion fraction exhibited by B3LYP-D4 is a direct consequence of the physical limitations inherent to the standard B3LYP functional, which lacks long-range dispersion and is inherently repulsive at intermediate van der Waals distances. When the semiclassical D4 correction is applied, it must first counteract this artificial electronic repulsion from the base functional before providing the actual binding attraction. This necessity for the D4 patch to overcompensate explains why the dispersion component alone can exceed the net binding energy ($E_{\textrm{disp}}/E_b > 1.0$). A similar, though less pronounced, trend is seen in PBE-D4 and PBE0-D4; standard GGAs and global hybrids lack an exact treatment of long-range electron correlation, forcing the explicit D4 correction to supply the bulk of the attractive force. PBE0-D4 exhibits a slightly lower dispersion fraction than PBE-D4 because the introduction of 25$\%$ exact Hartree-Fock exchange modifies the short-to-medium range Pauli repulsion profile, making the base functional marginally more competent near the binding equilibrium.

In contrast, the range-separated hybrids $\omega$B97X-D4rev and $\omega$B97M-V achieve a more physically rigorous and balanced description, in line with earlier reports \citep{mardirossianThirtyYearsDensity2017,goerigkLookDensityFunctional2017}. By incorporating a variable percentage of Hartree-Fock exchange that scales with electron distance (reaching 100$\%$ at long range), these functionals drastically mitigate self-interaction errors and provide a robust electronic structure foundation. Consequently, whether paired with the atom-pairwise D4 correction ($\omega$B97X-D4rev) or the non-local, density-dependent VV10 functional ($\omega$B97M-V), the base functional handles induction and electrostatics accurately, leading to highly consistent and physically realistic dispersion fractions. Finally, the anomalous behavior of M06-D3(0), which exhibits the lowest dispersion fraction across all configurations, is explained by its heavily parameterized meta-GGA framework. The M06 functional was explicitly optimized to capture medium-range correlation and dispersion-like attraction implicitly through its flexible kinetic energy density dependence. Because the base M06 functional already accounts for a substantial portion of the binding attraction at intermediate distances, the explicit, external D3 correction is only required to patch the remaining long-range asymptotic tail. This results in a very small explicit $E_{\textrm{disp}}$ value, leading to the exceptionally low dispersion fractions observed in Figure \ref{fig:BE100_functionals_dispersion}. It should in this context also be noted that in the reviews of Mardirossian and Goerigk, the M06-D3(0) is judged least suitable overall for the systems considered in this study, owing to the well-documented failure of Minnesota-type functionals to benefit consistently from D3-class dispersion corrections at non-equilibrium (e.g., adsorption-relevant) geometries and to their demonstrated oversensitivity to the integration grid, which manifests as unphysical oscillations in weakly bound potential energy curves \citep{mardirossianThirtyYearsDensity2017,goerigkLookDensityFunctional2017}.

\subsection{Astrophysical and Astrochemical Implications}
These findings have direct implications for astrochemical models of interstellar grain chemistry. A first implication of these results is their impact on snow lines and desorption temperatures. Astrochemical gas-grain models typically rely on binding energies to calculate thermal desorption rates via the Polanyi-Wigner equation. An error of 250 K or 500 K, which is typical when using water clusters that are too small ($N \le 20$), translates to an uncertainty of several Kelvin in the predicted desorption temperature of volatile species \citep{Anderl2016-jq}. In protoplanetary disks, a shift of several Kelvin in desorption temperature can move the calculated radial position of the snow lines of volatiles \citep{obergEFFECTSSNOWLINESPLANETARY2011}, even by up to several astronomical units (AU) for species with very low binding energies such as CO, directly affecting our understanding of volatile delivery to forming planetesimals.

Second, these results highlight the danger of single-value parameterisations in astrochemical networks. Indeed, most astrochemical networks assign a single, scalar binding energy to each molecular species. However, like many other recent calculations (see e.g. the studies mentioned in Table \ref{Table1} above), our results show a large standard deviation in binding energies across different ASW configurations (e.g., a spread of nearly $\pm 2000$ K for NH$_3$ on pristine surfaces). This structural heterogeneity means that on a real amorphous interstellar dust grain, molecules do not desorb at a single temperature. Instead, desorption occurs over a broad temperature range as molecules thermally migrate (at temperatures $\ge 20$ K) from weak pristine sites and become temporarily trapped in deeper structural cavities before finally escaping into the gas phase. Utilising average binding energies without accounting for this site-to-site distribution can lead to highly inaccurate gas-phase abundance predictions in cold molecular clouds.

Another modelling implication concerns the quantum chemical calculation of binding energies on ASW. In such calculations, computational chemists often focus heavily on the choice of density functional or basis set. However, our convergence results (Figure \ref{fig:convergence_functionals}) reveal that the error introduced by using a truncated cluster size (e.g., $N = 20$) is often 300 to 500 K. This cluster-size truncation error is of the same order or even larger than the typical energy differences obtained between different modern density functionals. Therefore, selecting an expensive, high-level functional is of little use if the underlying ice substrate model is too small to capture the long-range electrostatic environment. For reliable astrochemical data, computational studies must prioritise larger cluster sizes ($N \ge 40$ for flat surfaces, and perhaps $N \ge 80$ for defects/cavities) over highly expensive density functionals.

Finally, yet another important implication of these results concerns benchmarking against high-level quantum chemical techniques, such as CCSD(T). It is common practice in this field to select one or several density functionals based on a preceding benchmark of interaction energies against such a high-level technique, typically performed on very small ice clusters for computational tractability. However, our results show that the interaction energy strongly depends on cluster size. While Section \ref{subsec:func_indep} demonstrates that this size-dependence is remarkably consistent across six diverse density functionals, this consistency was established only within the Kohn-Sham DFT framework; it does not guarantee that a fundamentally different treatment of correlation and dispersion, such as CCSD(T), would exhibit the same size-dependence. Since there is no guarantee that the change in interaction energy as a function of cluster size is similar for the chosen DFT functional and the high-level reference method, the validity of benchmarking adsorption energies against very small clusters can be questioned.

\section{Conclusions}
This study systematically maps the convergence behaviour of molecular interaction energies on ASW cluster substrates as a function of size, providing a quantitative framework for computational astrochemical modelling. By assessing three relevant interstellar probes (CO, CO$_2$, and NH$_3$) across cluster sizes of $N = 10$ to 100 water molecules using six distinct density functionals, we isolated the spatial and electronic boundaries required to eliminate or at least reduce finite-size artefacts.

The investigation reveals a distinct decoupling between short-range and long-range non-covalent forces driving physisorption on interstellar ice. Due to the steep spatial decay of molecular van der Waals forces, the dispersion component stabilises rapidly and is virtually fully converged by $N = 40$ across all examined configurations. This confirms that dispersion is primarily modulated by the immediate, local coordination shells. Total interaction energies, in contrast, require significantly larger cluster sizes to achieve numerical stabilisation. The lingering drift observed at intermediate sizes is driven by the cluster's long-range electric field and cooperative matrix polarisation. Truncating clusters below $N = 30$ introduces severe unphysical fluctuations by abruptly severing these cooperative hydrogen-bonding networks.

Substrate morphology strongly dictates size requirements. While a cluster size of $N \approx 40$ H$_2$O molecules is sufficient to drop median convergence errors below the 100 K threshold on flat, pristine surfaces, complex structural cavities introduce highly variable polarisation pathways. Consequently, minimising worst-case convergence errors within these localised structural defects requires larger configurations ($N \ge 80$).  
The comparison of the six functionals indicates that the rate of size convergence is an intrinsic geometric and electrostatic property of the physical system rather than an artefact of the electronic structure method. While the absolute partitioning between dispersive and non-dispersive contributions varies widely by functional, the mean absolute deviation curves across all six methods are virtually superimposed.

The error introduced by utilising an inadequately truncated substrate model ($300-500$ K for $N \le 20$) consistently exceeds the nominal energy variations observed between different modern density functionals. Computational workflows must therefore prioritise sufficient cluster size ($N \ge 40$ for surfaces; $N \ge 80$ for cavities) over the deployment of computationally expensive functionals on insufficient substrate geometries.  

These findings carry immediate consequences for macroscopic astrochemical frameworks. Because thermal desorption rates scale exponentially with binding energy, the size-induced discrepancies quantified here prevent multi-astronomical-unit shifts in the modeled radial positions of disk snow lines, enhancing our ability to interpret high-resolution data from ALMA and JWST. Furthermore, the broad site-to-site energy distributions documented across different amorphous configurations, highlighted by a substantial spread for NH$_3$ on pristine sites, demonstrate the limitations of single-value scalar parameterisations.

Ultimately, moving toward cluster models that adequately encompass long-range polarisation networks, while transitioning kinetic models toward site-to-site energy distributions, will fundamentally improve the predictive reliability of gas-grain chemical networks in star-forming environments.

\begin{acknowledgments}
The resources and services used in this work were provided by the VSC (Flemish Supercomputer Centre), funded by the Research Foundation - Flanders (FWO) and the Flemish Government. This work was supported by the Marie Skłodowska–Curie Postdoctoral Fellowship (Grant Agreement No. 101211724).
\end{acknowledgments}

\begin{contribution}
ECN came up with the initial research concept, performed the calculations, analysed the data and wrote the manuscript.
IG, CK and TV performed ancillary calculations, participated in interpretation of the data and edited the manuscript.
ECN and IG obtained the funding that made this research possible.
\end{contribution}

\bibliography{Neyts_v2}{}

\begin{thebibliography}{}
\expandafter\ifx\csname natexlab\endcsname\relax\def\natexlab#1{#1}\fi
\providecommand{\url}[1]{\href{#1}{#1}}
\providecommand{\dodoi}[1]{doi:~\href{http://doi.org/#1}{\nolinkurl{#1}}}
\providecommand{\doeprint}[1]{\href{http://ascl.net/#1}{\nolinkurl{http://ascl.net/#1}}}
\providecommand{\doarXiv}[1]{\href{https://arxiv.org/abs/#1}{\nolinkurl{https://arxiv.org/abs/#1}}}

\bibitem[{S. Anderl {et~al.}(2016)Anderl, Maret, Cabrit, Belloche, Maury,
  Andr{\'e}, Codella, Bacmann, Bontemps, Podio, Gueth, \&
  Bergin}]{Anderl2016-jq}
Anderl, S., Maret, S., Cabrit, S., {et~al.} 2016, \bibinfo{title}{Probing the
  {CO} and methanol snow lines in young protostars,} Astronomy \& Astrophysics,
  591, A3, \dodoi{10.1051/0004-6361/201527831}

\bibitem[{F. Benedetti {et~al.}(2026)Benedetti, Satta, Grassi, Vogt-Geisse, \&
  Bovino}]{benedettiCODiffusionInterstellar2026}
Benedetti, F., Satta, M., Grassi, T., Vogt-Geisse, S., \& Bovino, S. 2026,
  \bibinfo{title}{CO {Diffusion} on {Interstellar} {Amorphous} {Solid} {Water}:
  A {Computational} {Study},} ACS Earth and Space Chemistry, 10, 224,
  \dodoi{10.1021/acsearthspacechem.5c00311}

\bibitem[{A. Boogert {et~al.}(2015)Boogert, Gerakines, \&
  Whittet}]{boogertObservationsIcyUniverse2015}
Boogert, A., Gerakines, P., \& Whittet, D. 2015, \bibinfo{title}{Observations
  of the {Icy} {Universe},} Annual Review of Astronomy and Astrophysics, 53,
  541, \dodoi{10.1146/annurev-astro-082214-122348}

\bibitem[{G.~M. Bovolenta {et~al.}(2025)Bovolenta, Molpeceres, Furuya, K{\"
  a}stner, \& Vogt-Geisse}]{bovolentaCOAdsorptionSites2025}
Bovolenta, G.~M., Molpeceres, G., Furuya, K., K{\" a}stner, J., \& Vogt-Geisse,
  S. 2025, \bibinfo{title}{CO adsorption sites on interstellar water ices
  explored with machine learning potentials: Binding energy distributions and
  snowline,} Astronomy \& Astrophysics, 703, A172,
  \dodoi{10.1051/0004-6361/202555836}

\bibitem[{G.~M. Bovolenta {et~al.}(2024)Bovolenta, Silva-Vera, Bovino,
  Molpeceres, K{\" a}stner, \&
  Vogt-Geisse}]{bovolentaInDepthExplorationCatalytic2024}
Bovolenta, G.~M., Silva-Vera, G., Bovino, S., {et~al.} 2024,
  \bibinfo{title}{In-{Depth} {Exploration} of {Catalytic} {Sites} on
  {Amorphous} {Solid} {Water}: I. {The} {Astrosynthesis} of {Aminomethanol},}
  Physical Chemistry Chemical Physics, 26, 18692, \dodoi{10.1039/D4CP01865F}

\bibitem[{G.~M. Bovolenta {et~al.}(2022)Bovolenta, Vogt-Geisse, Bovino, \&
  Grassi}]{bovolentaBindingEnergyEvaluation2022}
Bovolenta, G.~M., Vogt-Geisse, S., Bovino, S., \& Grassi, T. 2022,
  \bibinfo{title}{Binding {Energy} {Evaluation} {Platform}: A {Database} of
  {Quantum} {Chemical} {Binding} {Energy} {Distributions} for the
  {Astrochemical} {Community},} The Astrophysical Journal Supplement Series,
  262, 17, \dodoi{10.3847/1538-4365/ac7f31}

\bibitem[{S. Boys {\&} F. Bernardi(1970)Boys \&
  Bernardi}]{boysCalculationSmallMolecular1970}
Boys, S., \& Bernardi, F. 1970, \bibinfo{title}{The calculation of small
  molecular interactions by the differences of separate total energies. {Some}
  procedures with reduced errors,} Molecular Physics, 19, 553,
  \dodoi{10.1080/00268977000101561}

\bibitem[{E. Caldeweyher {et~al.}(2017)Caldeweyher, Bannwarth, \&
  Grimme}]{caldeweyherExtensionD3Dispersion2017}
Caldeweyher, E., Bannwarth, C., \& Grimme, S. 2017, \bibinfo{title}{Extension
  of the {D3} dispersion coefficient model,} The Journal of Chemical Physics,
  147, 034112, \dodoi{10.1063/1.4993215}

\bibitem[{M.~P. Collings {et~al.}(2004)Collings, Anderson, Chen, Dever, Viti,
  Williams, \& McCoustra}]{collingsLaboratorySurveyThermal2004}
Collings, M.~P., Anderson, M.~A., Chen, R., {et~al.} 2004, \bibinfo{title}{A
  laboratory survey of the thermal desorption of astrophysically relevant
  molecules,} Monthly Notices of the Royal Astronomical Society, 354, 1133,
  \dodoi{10.1111/j.1365-2966.2004.08272.x}

\bibitem[{H.~M. Cuppen {et~al.}(2024)Cuppen, Linnartz, \&
  Ioppolo}]{cuppenLaboratoryComputationalStudies2024}
Cuppen, H.~M., Linnartz, H., \& Ioppolo, S. 2024, \bibinfo{title}{Laboratory
  and {Computational} {Studies} of {Interstellar} {Ices},} Annual Review of
  Astronomy and Astrophysics, 62, 243,
  \dodoi{10.1146/annurev-astro-071221-052732}

\bibitem[{H.~M. Cuppen {et~al.}(2017)Cuppen, Walsh, Lamberts, Semenov, Garrod,
  Penteado, \& Ioppolo}]{cuppenGrainSurfaceModels2017}
Cuppen, H.~M., Walsh, C., Lamberts, T., {et~al.} 2017, \bibinfo{title}{Grain
  {Surface} {Models} and {Data} for {Astrochemistry},} Space Science Reviews,
  212, 1, \dodoi{10.1007/s11214-016-0319-3}

\bibitem[{G. Di~Genova {et~al.}(2025)Di~Genova, Perrero, Rosi, Ceccarelli,
  Rimola, \& Balucani}]{digenovaHotSulfurRocks2025}
Di~Genova, G., Perrero, J., Rosi, M., {et~al.} 2025, \bibinfo{title}{Hot
  {Sulfur} on the {Rocks}: The {Reaction} of {Electronically} {Excited}
  {Sulfur} {Atoms} with {Water} in an {Ice}-{Surface} {Model},} ACS Earth and
  Space Chemistry, 9, 844, \dodoi{10.1021/acsearthspacechem.4c00351}

\bibitem[{D. Duflot {et~al.}(2021)Duflot, Toubin, \&
  Monnerville}]{duflotTheoreticalDeterminationBinding2021}
Duflot, D., Toubin, C., \& Monnerville, M. 2021, \bibinfo{title}{Theoretical
  {Determination} of {Binding} {Energies} of {Small} {Molecules} on
  {Interstellar} {Ice} {Surfaces},} Frontiers in Astronomy and Space Sciences,
  8, 645243, \dodoi{10.3389/fspas.2021.645243}

\bibitem[{J. Enrique-Romero {et~al.}(2022)Enrique-Romero, Rimola, {Ceccarelli},
  Ugliengo, Balucani, \& Skouteris}]{Enrique2022Quantum}
Enrique-Romero, J., Rimola, A., {Ceccarelli}, {et~al.} 2022,
  \bibinfo{title}{Quantum mechanical simulations of the radical-radical
  chemistry on icy surfaces,} The Astrophysical Journal Supplement Series, 259,
  39, \dodoi{10.3847/1538-4365/ac480e}

\bibitem[{S. Ferrero {et~al.}(2020)Ferrero, Zamirri, Ceccarelli, Witzel,
  Rimola, \& Ugliengo}]{ferreroBindingEnergiesInterstellar2020}
Ferrero, S., Zamirri, L., Ceccarelli, C., {et~al.} 2020,
  \bibinfo{title}{Binding energies of interstellar molecules on crystalline and
  amorphous models of water ice by ab-initio calculations,} The Astrophysical
  Journal, 904, 11, \dodoi{10.3847/1538-4357/abb953}

\bibitem[{H.~J. Fraser {et~al.}(2001)Fraser, Collings, McCoustra, \&
  Williams}]{fraserThermalDesorptionWater2001}
Fraser, H.~J., Collings, M.~P., McCoustra, M. R.~S., \& Williams, D.~A. 2001,
  \bibinfo{title}{Thermal desorption of water ice in the interstellar medium,}
  Monthly Notices of the Royal Astronomical Society, 327, 1165,
  \dodoi{10.1046/j.1365-8711.2001.04835.x}

\bibitem[{L. Goerigk {et~al.}(2017)Goerigk, Hansen, Bauer, Ehrlich, Najibi, \&
  Grimme}]{goerigkLookDensityFunctional2017}
Goerigk, L., Hansen, A., Bauer, C., {et~al.} 2017, \bibinfo{title}{A look at
  the density functional theory zoo with the advanced {GMTKN55} database for
  general main group thermochemistry, kinetics and noncovalent interactions,}
  Physical Chemistry Chemical Physics, 19, 32184, \dodoi{10.1039/C7CP04913G}

\bibitem[{S. Grimme(2011)Grimme}]{grimmeDensityFunctionalTheory2011}
Grimme, S. 2011, \bibinfo{title}{Density functional theory with {London}
  dispersion corrections,} WIREs Computational Molecular Science, 1, 211,
  \dodoi{10.1002/wcms.30}

\bibitem[{M. Groyne {et~al.}(2025)Groyne, Champagne, Baijot, \&
  De~Becker}]{groyneRobustBindingEnergy2025}
Groyne, M., Champagne, B., Baijot, C., \& De~Becker, M. 2025,
  \bibinfo{title}{Robust binding energy distribution sampling on amorphous
  solid water models: Method testing and validation with {NH}\textsubscript{3}
  , {CO}, and {CH}\textsubscript{4},} Astronomy \& Astrophysics, 698, A284,
  \dodoi{10.1051/0004-6361/202555097}

\bibitem[{T.~I. Hasegawa {et~al.}(1992)Hasegawa, Herbst, \&
  Leukng}]{hasegawaModelsGasgrainChemistry1992}
Hasegawa, T.~I., Herbst, E., \& Leukng, C.~M. 1992, \bibinfo{title}{Models of
  gas-grain chemistry in dense interstellar clouds with complex organic
  molecules,} The Astrophysical Journal Supplement Series, 82, 167,
  \dodoi{10.1086/191713}

\bibitem[{B. Helmich-Paris {et~al.}(2021)Helmich-Paris, De~Souza, Neese, \&
  Izs{\' a}k}]{Helmich2021improved}
Helmich-Paris, B., De~Souza, B., Neese, F., \& Izs{\' a}k, R. 2021,
  \bibinfo{title}{An improved chain of spheres for exchange algorithm,} The
  Journal of Chemical Physics, 155, 104109, \dodoi{10.1063/5.0058766}

\bibitem[{L. Hornek\ae{}r {et~al.}(2005)Hornek\ae{}r, Baurichter, Petrunin,
  Luntz, Kay, \& Al-Halabi}]{hornekaerInfluenceSurfaceMorphology2005}
Hornek\ae{}r, L., Baurichter, A., Petrunin, V.~V., {et~al.} 2005,
  \bibinfo{title}{Influence of surface morphology on {D2} desorption kinetics
  from amorphous solid water,} The Journal of Chemical Physics, 122, 124701,
  \dodoi{10.1063/1.1874934}

\bibitem[{H. Kakkar {et~al.}(2025)Kakkar, Mart{\' i}nez-Bachs, Ceccarelli,
  Ugliengo, \& Rimola}]{kakkarBindingEnergiesInterstellar2025}
Kakkar, H., Mart{\' i}nez-Bachs, B., Ceccarelli, C., Ugliengo, P., \& Rimola,
  A. 2025, \bibinfo{title}{Binding {Energies} of {Interstellar} {Complex}
  {Organic} {Molecules} on {Water} {Ice} {Surfaces}: A {Quantum} {Chemical}
  {Investigation},} The Astrophysical Journal, 993, 184,
  \dodoi{10.3847/1538-4357/ae064a}

\bibitem[{T.
  Lamberts(2018)Lamberts}]{lambertsInterstellarCarbonMonosulfide2018}
Lamberts, T. 2018, \bibinfo{title}{From interstellar carbon monosulfide to
  methyl mercaptan: paths of least resistance,} Astronomy \& Astrophysics, 615,
  L2, \dodoi{10.1051/0004-6361/201832830}

\bibitem[{N. Mardirossian {\&} M. Head-Gordon(2017)Mardirossian \&
  Head-Gordon}]{mardirossianThirtyYearsDensity2017}
Mardirossian, N., \& Head-Gordon, M. 2017, \bibinfo{title}{Thirty years of
  density functional theory in computational chemistry: an overview and
  extensive assessment of 200 density functionals,} Molecular Physics, 115,
  2315, \dodoi{10.1080/00268976.2017.1333644}

\bibitem[{L. Mart{\' i}nez {et~al.}(2009)Mart{\' i}nez, Andrade, Birgin, \&
  Mart{\' i}nez}]{martinezPACKMOLPackageBuilding2009}
Mart{\' i}nez, L., Andrade, R., Birgin, E.~G., \& Mart{\' i}nez, J.~M. 2009,
  \bibinfo{title}{P \textsc{{ACKMOL}} : A package for building initial
  configurations for molecular dynamics simulations,} Journal of Computational
  Chemistry, 30, 2157, \dodoi{10.1002/jcc.21224}

\bibitem[{M.~K. McClure {et~al.}(2023)McClure, Rocha, Pontoppidan, Crouzet,
  Chu, Dartois, Lamberts, Noble, Pendleton, Perotti, Qasim, Rachid, Smith, Sun,
  Beck, Boogert, Brown, Caselli, Charnley, Cuppen, Dickinson, Drozdovskaya,
  Egami, Erkal, Fraser, Garrod, Harsono, Ioppolo, Jimenez-Serra, Jin,
  J\o{}rgensen, Kristensen, Lis, McCoustra, McGuire, Melnick, Oberg, Palumbo,
  Shimonishi, Sturm, Dishoeck, \& Linnartz}]{mcclureIceAgeJWST2023}
McClure, M.~K., Rocha, W. R.~M., Pontoppidan, K.~M., {et~al.} 2023,
  \bibinfo{title}{An {Ice} {Age} {JWST} inventory of dense molecular cloud
  ices,} Nature Astronomy, 7, 431, \dodoi{10.1038/s41550-022-01875-w}

\bibitem[{G. Molpeceres {et~al.}(2023)Molpeceres, Enrique-Romero, \&
  Aikawa}]{molpeceresCrackingPuzzleCO22023}
Molpeceres, G., Enrique-Romero, J., \& Aikawa, Y. 2023,
  \bibinfo{title}{Cracking the puzzle of {CO}\textsubscript{2} formation on
  interstellar ices: Quantum chemical and kinetic study of the {CO} + {OH}
  \textrightarrow{} {CO}\textsubscript{2} + {H} reaction,} Astronomy \&
  Astrophysics, 677, A39, \dodoi{10.1051/0004-6361/202347097}

\bibitem[{A.~A. Mozhegorov {\&} A.~I. Vasyunin(2026)Mozhegorov \&
  Vasyunin}]{mozhegorovAtomAmorphousH22026}
Mozhegorov, A.~A., \& Vasyunin, A.~I. 2026, \bibinfo{title}{H {Atom} {Near} the
  {Amorphous} {H}\textsubscript{2} {O} {Surface}: The {Revision} of {Binding}
  {Energy} {Estimations},} ACS Earth and Space Chemistry, 10, 982,
  \dodoi{10.1021/acsearthspacechem.5c00270}

\bibitem[{F. Neese(2023)Neese}]{neeseSHARKIntegralGeneration2023}
Neese, F. 2023, \bibinfo{title}{The \textsc{{SHARK}} integral generation and
  digestion system,} Journal of Computational Chemistry, 44, 381,
  \dodoi{10.1002/jcc.26942}

\bibitem[{F. Neese(2025)Neese}]{neeseSoftwareUpdateORCA2025}
Neese, F. 2025, \bibinfo{title}{Software {Update}: The \textsc{{ORCA}}
  {Program} {System}---{Version} 6.0,} WIREs Computational Molecular Science,
  15, e70019, \dodoi{10.1002/wcms.70019}

\bibitem[{J.~A. Noble {et~al.}(2012)Noble, Congiu, Dulieu, \&
  Fraser}]{nobleThermalDesorptionCharacteristics2012}
Noble, J.~A., Congiu, E., Dulieu, F., \& Fraser, H.~J. 2012,
  \bibinfo{title}{Thermal desorption characteristics of {CO}, {O2} and {CO2} on
  non-porous water, crystalline water and silicate surfaces at submonolayer and
  multilayer coverages: Desorption from {H2O}(np), {H2O}(c) and {SiOx},}
  Monthly Notices of the Royal Astronomical Society, 421, 768,
  \dodoi{10.1111/j.1365-2966.2011.20351.x}

\bibitem[{K.~I. {\" O}berg {et~al.}(2011){\" O}berg, Murray-Clay, \&
  Bergin}]{obergEFFECTSSNOWLINESPLANETARY2011}
{\" O}berg, K.~I., Murray-Clay, R., \& Bergin, E.~A. 2011, \bibinfo{title}{THE
  {EFFECTS} {OF} {SNOWLINES} {ON} {C}/{O} {IN} {PLANETARY} {ATMOSPHERES},} The
  Astrophysical Journal, 743, L16, \dodoi{10.1088/2041-8205/743/1/L16}

\bibitem[{J. Perrero {et~al.}(2022)Perrero, Enrique-Romero, Ferrero,
  Ceccarelli, Podio, Codella, Rimola, \&
  Ugliengo}]{perreroBindingEnergiesInterstellar2022}
Perrero, J., Enrique-Romero, J., Ferrero, S., {et~al.} 2022,
  \bibinfo{title}{Binding {Energies} of {Interstellar} {Relevant} {S}-bearing
  {Species} on {Water} {Ice} {Mantles}: A {Quantum} {Mechanical}
  {Investigation},} The Astrophysical Journal, 938, 158,
  \dodoi{10.3847/1538-4357/ac9278}

\bibitem[{J. Perrero {et~al.}(2024)Perrero, Vitorino, Congiu, Ugliengo, Rimola,
  \& Dulieu}]{perreroBindingEnergiesEthanol2024}
Perrero, J., Vitorino, J., Congiu, E., {et~al.} 2024, \bibinfo{title}{Binding
  energies of ethanol and ethylamine on interstellar water ices: synergy
  between theory and experiments,} Physical Chemistry Chemical Physics, 26,
  18205, \dodoi{10.1039/D4CP01934B}

\bibitem[{E.~L. Piacentino {\&} K.~I. {\" O}berg(2022)Piacentino \& {\"
  O}berg}]{piacentinoComputationalEstimationBinding2022}
Piacentino, E.~L., \& {\" O}berg, K.~I. 2022, \bibinfo{title}{Computational
  {Estimation} of the {Binding} {Energies} of {PO}\textsubscript{x} and
  {HPO}\textsubscript{x} (x = 2, 3) {Species},} The Astrophysical Journal, 939,
  93, \dodoi{10.3847/1538-4357/ac96e2}

\bibitem[{C. Qi {et~al.}(2013)Qi, {\" O}berg, Wilner,
  D\textquoteright{}Alessio, Bergin, Andrews, Blake, Hogerheijde, \&
  Van~Dishoeck}]{qiImagingCOSnow2013a}
Qi, C., {\" O}berg, K.~I., Wilner, D.~J., {et~al.} 2013,
  \bibinfo{title}{Imaging of the {CO} {Snow} {Line} in a {Solar} {Nebula}
  {Analog},} Science, 341, 630, \dodoi{10.1126/science.1239560}

\bibitem[{A. Rimola {et~al.}(2021)Rimola, Ceccarelli, Balucani, \&
  Ugliengo}]{rimolaInteractionHCOCations2021}
Rimola, A., Ceccarelli, C., Balucani, N., \& Ugliengo, P. 2021,
  \bibinfo{title}{Interaction of {HCO}+ {Cations} {With} {Interstellar}
  {Negative} {Grains}. {Quantum} {Chemical} {Investigation} and {Astrophysical}
  {Implications},} Frontiers in Astronomy and Space Sciences, 8, 655405,
  \dodoi{10.3389/fspas.2021.655405}

\bibitem[{W.~M.~C. Sameera {et~al.}(2021{\natexlab{a}})Sameera, Senevirathne,
  Andersson, Al-lbadi, Hidaka, Kouchi, Nyman, \&
  Watanabe}]{sameeraCH3RadicalBinding2021}
Sameera, W. M.~C., Senevirathne, B., Andersson, S., {et~al.}
  2021{\natexlab{a}}, \bibinfo{title}{CH\textsubscript{3} {O} {Radical}
  {Binding} on {Hexagonal} {Water} {Ice} and {Amorphous} {Solid} {Water},} The
  Journal of Physical Chemistry A, 125, 387, \dodoi{10.1021/acs.jpca.0c09111}

\bibitem[{M. Sil {et~al.}(2024)Sil, Roy, Gorai, Nakatani, Shimonishi, Furuya,
  Inostroza-Pino, Caselli, \& Das}]{silAssessingRealisticBinding2024}
Sil, M., Roy, A., Gorai, P., {et~al.} 2024, \bibinfo{title}{Assessing realistic
  binding energies of some essential interstellar radicals with amorphous solid
  water: A fully quantum chemical approach,} Astronomy \& Astrophysics, 690,
  A252, \dodoi{10.1051/0004-6361/202451642}

\bibitem[{N. Tieppo {et~al.}(2025)Tieppo, Redondo, Pauzat, Parisel, Guillemin,
  \& Ellinger}]{tieppoBuildingFormamideNsubstituted2025}
Tieppo, N., Redondo, P., Pauzat, F., {et~al.} 2025, \bibinfo{title}{Building
  formamide and {N}-substituted formamides from isocyanates on hydrogenated
  water ices,} Astronomy \& Astrophysics, 695, A133,
  \dodoi{10.1051/0004-6361/202450614}

\bibitem[{T. Vorsselmans {et~al.}(2026)Vorsselmans, Grubova, Verhagen,
  Guldentops, Buimer, King, \& Neyts}]{vorsselmansEffectIceCharging2026}
Vorsselmans, T., Grubova, I., Verhagen, K., {et~al.} 2026,
  \bibinfo{title}{Effect of ice charging on the astrochemistry of interstellar
  sulfur-bearing species on amorphous solid water,} Astronomy \& Astrophysics,
  \dodoi{10.1051/0004-6361/202661939}

\bibitem[{T. Vorsselmans {\&} E.~C. Neyts(2025)Vorsselmans \&
  Neyts}]{vorsselmansBindingEnergiesSmall2025b}
Vorsselmans, T., \& Neyts, E.~C. 2025, \bibinfo{title}{Binding {Energies} of
  {Small} {Interstellar} {Molecules} on {Neutral} and {Charged} {Amorphous}
  {Solid} {Water} {Surfaces},} The Astrophysical Journal, 986, 30,
  \dodoi{10.3847/1538-4357/add145}

\bibitem[{L. Zamirri {et~al.}(2019)Zamirri, Ugliengo, Ceccarelli, \&
  Rimola}]{zamirriQuantumMechanicalInvestigations2019}
Zamirri, L., Ugliengo, P., Ceccarelli, C., \& Rimola, A. 2019,
  \bibinfo{title}{Quantum {Mechanical} {Investigations} on the {Formation} of
  {Complex} {Organic} {Molecules} on {Interstellar} {Ice} {Mantles}. {Review}
  and {Perspectives},} ACS Earth and Space Chemistry, 3, 1499,
  \dodoi{10.1021/acsearthspacechem.9b00082}

\end{thebibliography}
\bibliographystyle{aasjournalv7.1}

\end{document}